\documentclass[rnote]{aa}  

\usepackage{graphicx}
\usepackage{natbib}
\bibpunct{(}{)}{;}{a}{}{,} 

\usepackage{txfonts}
\begin{document}

\title{Vindicating single-$T$ modified blackbody fits to Herschel 
SEDs\thanks{{\it Herschel} is an ESA space observatory with science 
instruments provided by European-led Principal Investigator consortia 
and with important participation from NASA.}}
\author{Simone Bianchi}
\institute{INAF-Osservatorio Astrofisico di Arcetri, Largo E. Fermi 5, I-50125, Florence, Italy}
\date{Received ; accepted }

 
\abstract{
I show here that the bulk of the dust mass in a galaxy can be equivalently 
estimated from: i) the full spectral energy distribution of dust emission, 
using the approach of \citet{DraineApJ2007b} that includes a distribution 
of dust grains and a range of interstellar radiation field intensities;
ii) the emission in the wavelength range $100\mu m\le\lambda\le500\mu m$ 
(covered by the Herschel Space Observatory), by fitting to the data
a simpler single temperature modified blackbody.
Recent claims on the contrary \citep{DaleApJ2012} should be interpreted 
as a caveat to use in the simpler fits an absorption cross section
which is consistent both in the normalization and in the spectral index 
$\beta$ with that of the full dust model. I also show that the dust mass
does not depend significantly on the choice of $\beta$, if both the dust
mass and the absorption cross section are derived with the same assumption 
on $\beta$.
}

\keywords{
dust, extinction - radiation mechanisms: thermal - infrared, submillimeter: ISM, galaxies 
}

   \maketitle
%

\section{Introduction}

A full coverage of the Spectral Energy Distribution (SED) of dust emission in 
galaxies has recently become available, thanks mainly to the Spitzer Space 
Telescope \citep{WernerApJS2004} and to the Herschel Space Observatory 
\citep{PilbrattA&A2010}. Dust emission models can now 
extract from the infrared observations a wealth of information on the dust 
composition, grain sizes and intensity of the dust heating sources. One such 
model is that of \citet[hereafter, DL07]{DraineApJ2007b}, which has been successful 
in reproducing both the global \citep[hereafter, D12]{DraineApJ2007,DaleApJ2012} 
and the resolved SEDs \citep{AnianoApJ2012}.

A general result of the modelling is that emission for $\lambda \ge 100\mu$m
predominantly comes from dust heated at thermal equilibrium by a mean 
interstellar radiation field (ISRF); and that this dust component 
constitutes the bulk of the dust mass in a galaxy
\citep[$\approx$98-99\%;][D12]{DraineApJ2007}. 
If all dust grains share the same size and composition, this emission is 
equivalent to that of a single temperature modified blackbody (MBB), i.e.\
a blackbody multiplied by the dust absorption cross section. 
For a dust model including grains of different sizes and compositions, 
and thus different absorption cross sections, the SED could be broader than that of
a MBB, because different grains attain a range of thermal equilibrium 
temperatures. Thus, the mass obtained by fitting to the observed SED 
a MBB with an average absorption cross section could in principle be biased 
with respect to that derived with the DL07 approach, using the full dust grain model.
Nevertheless \citet{MagriniA&A2011},
fitting Herschel data at $\lambda \ge 100\mu$m for a sample of Virgo 
Cluster galaxies from the HeViCS programme \citep{DaviesMNRAS2012}, 
found that MBB dust masses are within $\approx$10\% of DL07 masses,
proving that the SED broadening due to the individual grain temperatures is minimal.

Using the same spectral range for the SEDs of galaxies in the Herschel
KINGFISH sample \citep{KennicuttPASP2011}, D12 instead claimed
that the MBB approach can underestimate the dust mass by up to a factor two, 
because "single blackbody curves do not capture the full range of dust temperatures 
inherent to any galaxy".

Starting from the D12 dataset, in this note I derive independently 
the MBB dust masses. I confirm the results of \citet{MagriniA&A2011}
(Sect.~\ref{sec_mass}). I also show that a meaningful derivation of the dust
mass can be obtained only if the same spectral index is used both in the MBB 
fitting and in the derivation of the absorption cross section (Sect.~\ref{sec_beta}).
Conclusions are drawn in Sect.~\ref{sec_conclu}.

\begin{figure*}
\centering
\includegraphics[width=\hsize]{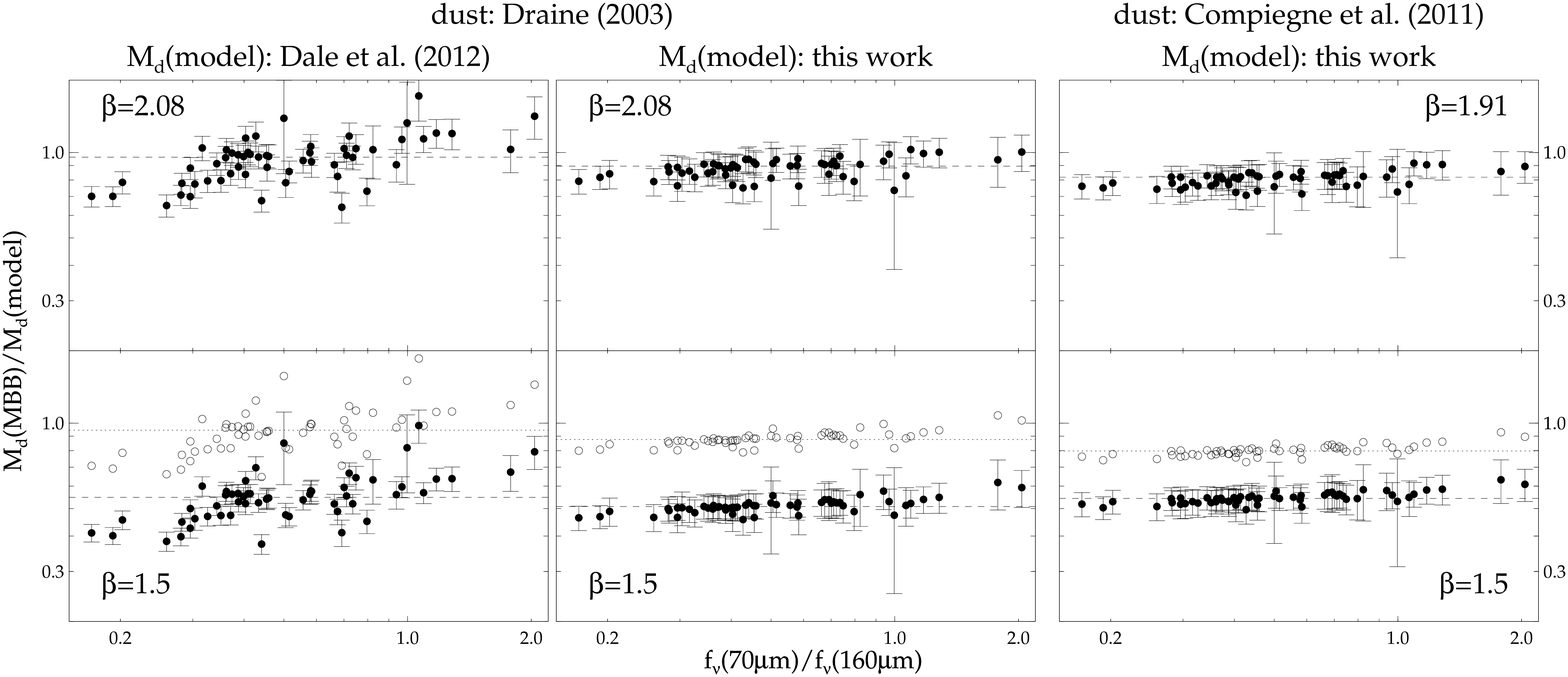}
\caption{
Ratio between dust masses derived from MBB fits to the 100-500$\mu$m SEDs 
of KINGFISH galaxies, M$_\mathrm{d}$(MBB), and those obtained using a dust emission 
model, M$_\mathrm{d}$(model). In the left panels, M$_\mathrm{d}$(model) is
obtained by D12 applying the DL07 method to the whole dust SED.
In the central panels, M$_\mathrm{d}$(model) comes from a simplified application 
of the DL07 approach to just the Herschel 100-500$\mu$m SED; the same
in the right panels, but using the dust properties from C11 (see
text for details).
In the top panels, M$_\mathrm{d}$(MBB) is derived using a power-law 
fit to the dust absorption cross section from D03 (left and 
center) and from C11 (right). In the bottom panels, the
spectral index is changed from the fitted value to $\beta=1.5$.
Dashed lines show the median value of the ratios. Open circles 
(and the dotted line) refer to the case for $\beta=1.5$ after correcting the 
$\kappa_\mathrm{abs}$ normalization (see Sect.~\ref{sec_beta} for details).
}
\label{fig_ratio}
\end{figure*}

\section{Dust masses from modified blackbody fits}
\label{sec_mass}

D12 measured the flux densities of the KINGFISH galaxies
observed by Herschel with the PACS instrument at 70, 100 and 160$\mu$m \citep{PoglitschA&A2010},
and the SPIRE instrument at 250, 350 and 500$\mu$m \citep{GriffinA&A2010}. After complementing the
Herschel observations with available infrared observations at shorter wavelengths, they
derived various parameters characterizing the dust emission, including the dust
mass, by using the DL07 approach. Here I use their photometry 
for the five Herschel bands with $\lambda\ge100\mu$m, and derive the MBB dust
mass for the objects detected in all five bands (56 objects out of 61).
The distances of the galaxies is taken from \citet{KennicuttPASP2011}.

Under the assumption that all dust grains share a single temperature $T_\mathrm{d}$,
and that the dust distribution is optically thin, the mass of dust $M_\mathrm{d}$ can be 
estimated by fitting the observed flux densities $f_\nu$ to a MBB,
\begin{equation}
f_\nu=\frac{M_\mathrm{d}}{D^2} \; \kappa_\mathrm{abs} \; B_\nu(T_\mathrm{d}),
\label{eq_mbb}
\end{equation}
where $D$ is the distance of the object, $B_\nu(T_\mathrm{d})$ the Planck function,
and $\kappa_\mathrm{abs}$ the grain absorption cross section per unit mass (a quantity 
sometimes referred to as {\em emissivity}). A power law was assumed for $\kappa_\mathrm{abs}$,
\begin{equation}
\kappa_\mathrm{abs} = \kappa_\mathrm{abs} (\lambda_0) \times 
\left(\frac{\lambda_0}{\lambda}\right)^\beta.
\label{eq_emy}
\end{equation}
Fits were performed using the procedure of \citet{MagriniA&A2011}: since the flux 
densities of D12 are not color-corrected, the MBB was integrated over the the PACS 
and SPIRE filter response functions before comparing it to the data; for SPIRE, the 
filter response functions for extended sources was selected.
The MPFIT IDL $\chi^2$ minimization routines were used \citep{MarkwardtProc2009}.
Recently, concern has been raised on the use of standard $\chi^2$ minimizations
techniques when fitting MBBs to observed SED, with biases arising when both 
$T_\mathrm{d}$ and $\beta$ are derived, resulting in a spurious
$T_\mathrm{d}$-$\beta$  anticorrelation that might (or might not) conceal real
grain properties \citep[see e.g.\ ][]{KellyApJ2012}.
However, in this work I always keep $\beta$ fixed and only derive $M_\mathrm{d}$ and 
$T_\mathrm{d}$, in analogy to the other works I compare with. Results from the
fit where checked with a bootstrapping technique, by fitting one hundred random 
representations of each SED (each compatible with the original photometry, within 
its error). The mean and standard deviations for $M_\mathrm{d}$ and  $T_\mathrm{d}$ 
obtained with the bootstrapping were almost the same as those obtained by fitting
the observed SED with MPFIT. For the photometric errors provided by D12, the
relative errors obtained by the fit on M$_\mathrm{d}$ and $T_\mathrm{d}$ are, 
on average, 8 and 2\%, respectively.

Since the DL07 emission model used by D12 is based on \citeauthor{DraineARA&A2003}'s 
(\citeyear{DraineARA&A2003}; hereafter, D03)
model for Milky Way (MW) dust , an appropriate choice for $\kappa_\mathrm{abs}$ is the absorption 
cross section of the latter, averaged over the the grain size distribution and composition. 
The $R_\mathrm{V}=3.1$ MW dust model\footnote{The
averaged absorption cross section for this model is available at:
http://www.astro.princeton.edu/~draine/dust/dustmix.html.
Further updates to the model parameters \citep[DL07;][]{AnianoApJ2012} 
did not change substantially the mean dust absorption cross section for 
the wavelength range considered here (Draine, private communication).}
absorption cross section for $70\mu m<\lambda<700\mu m$ is well fitted
by Eq.~\ref{eq_emy} with
\begin{equation}
\kappa_\mathrm{abs} (250\mu\mathrm{m}) =4.0 \,\, \mathrm{cm^2\,\,  g^{-1}},\,\,\,\,\,\, \beta=2.08.
\label{fit_d03}
\end{equation}

The ratio between the dust masses obtained with the two methods is shown in 
Fig.~\ref{fig_ratio} (upper-left panel). As in the analogous Fig.~9 of D12, 
the ratio is plotted versus the 70$\mu$m-to-160$\mu$m flux density ratio. 
The dust mass obtained by fitting a MBB to the $100\mu m\le\lambda\le500\mu m$
data is quite close to that obtained with the DL07 model. The median
ratio is 0.96\footnote{In a number of Herschel papers 
\citep[see e.g.\ ][]{MagriniA&A2011,DaviesMNRAS2012,SmithApJ2012},
it was used
\[
\kappa_\mathrm{abs} (350\mu\mathrm{m}) =1.92 \,\, \mathrm{cm^2\,\,  g^{-1}},\,\,\,\,\,\, \beta=2,
\]
where the normalization at $350 \mu$m is the value for
the $R_\mathrm{V}=3.1$ Milky Way (MW) averaged dust absorption cross section from D03.
The median ratio obtained using these values is 0.95, almost the same of that
obtained using the fit of Eq.~\ref{fit_d03}. When the exact, tabulated, 
averaged cross section from D03's model is used, the
ratio becomes 0.98. It is to be noted that I consider here the full dust mass 
obtained by D12. If, in addition to the use of the tabulated 
values, I correct for the fraction of dust that is 
heated by higher intensity radiation fields (as given by the $\gamma$ parameter 
provided by D12) and does not contribute significantly to the 
emission at $\lambda \ge 100\mu$m, the median ratio rises to 1.0.}.

D12 claimed that the dust masses derived with a $\beta=2$ MBB
fit are underestimated, on average, by 25\% (though it appears that they have 
included the 70$\mu$m flux density in the fits, which I don't use here).
However, they adopted $\kappa_\mathrm{abs}(250\mu\mathrm{m})=4.8\,\mathrm{cm^2\, g^{-1}}$,
a value 20\% higher than that of the more recent D03's model
(i.e.\ they underestimated the MBB masses by 20\%). If they had used a cross 
section consistent with the dust model used within their implementation of the 
DL07 method, their result would have been in line with what I find here.

The dispersion of the mass ratios is 0.18. As it can be seen from 
Fig.~\ref{fig_ratio}, this is higher than the error on the ratio, whose mean is 
0.12 (estimated using the error from the fits for individual MBB masses, and 
assuming the same errors for the DL07 masses, which were not provided). 
The large scatter is 
related to the intrinsic differences of the two methods: the masses from
D12 were derived using additional fluxes, not just the five 
Herschel datapoints, with emission from dust heated by the mean ISRF 
constrained by shorter wavelength observations also; the dust
absorption cross section also depend, though slightly for the Herschel wavelength 
range, on the fraction of the dust mass in the form of polycyclic aromatic 
hydrocarbons (PAHs), the parameter $q_\mathrm{PAH}$ which is a result of
D12 fits. 

Indeed if these differences are removed, the scatter is reduced.  This was tested 
deriving DL07 masses using a simplified approach: I have used the emission 
models\footnote{The original DL07 models are available at: 
http://www.astro.princeton.edu/~draine/dust/irem.html.}
for dust heated by a single intensity of the ISRF only (as defined by the $U$ 
parameter in DL07, a scale factor with respect to the local Galactic ISRF from 
\citealt{MathisA&A1983}); and those computed for the $R_\mathrm{V}=3.1$ MW dust 
model only (i.e.\ $q_\mathrm{PAH}=4.68$\%; DL07).
As in the full DL07 approach, emission templates for different intensites were 
convolved with the filter response functions, but only for the five Herschel 
datapoints considered for the MBB fits. SED fits were produced by minimizing 
$\chi^2$ over the provided $U$ grid and over $M_\mathrm{d}$, and errors on the 
two quantities were computed with the same bootstrapping technique used for MBBs.
This confirmed that the errors on $M_\mathrm{d}$ are similar for both the MBB and
DL07 approach, as assumed earlier in this section.
The results of the fits are compatible with those of the full analysis in D12, 
though the $U$s found here are generally lower than their equivalent $U_\mathrm{min}$ 
(a distribution of $U$s is used in the full approach, the dust heated by 
$U_\mathrm{min}$ being responsible for most of the FIR peak). As a consequence, the 
DL07 masses derived here are larger, but only by $\sim 10$\%, the median ratio
is smaller, 0.9, and the scatter is reduced to 0.07, of the same order of the 
error (Fig.~\ref{fig_ratio}, upper-middle panel; the ratio is computed
using the same MBB masses of top-left panel). This result is identical to the tests 
we did in \citet{MagriniA&A2011}. 

The DL07 approach can be used with any dust grain model. As a further test,
I use it here with the MW dust model by \citet[hereafter, C11]{CompiegneA&A2011}, which
basically differs from D03's model in the use of
optical properties for amorphous carbon. Using the DustEM code described in
that paper\footnote{Available at: http://www.ias.u-psud.fr/DUSTEM/}
I have computed the emission for a grid of $U$ values, and derived the dust
masses as described in the previous paragraph.  DustEM also provides the
mean absorption cross section of the dust distribution, which can be fitted
with Eq.~\ref{eq_emy} and
\begin{equation}
\kappa_\mathrm{abs} (250\mu\mathrm{m}) =5.1 \,\, \mathrm{cm^2\,\,  g^{-1}},\,\,\,\,\,\, \beta=1.91.
\label{fit_mc10}
\end{equation}
As a results of the different $\kappa_\mathrm{abs}$, dust masses 
derived using the simplified, DL07 approach with the C11 model 
are a factor 0.73 the corresponding masses using D03.
When MBB masses are computed using Eq.~\ref{fit_mc10}, the median
ratio between MBB and DL07 masses for the C11 model
is 0.82, with a scatter of 0.05\footnote{It raises to 0.84 if the tabulated 
absorption cross section is used.} (Fig.~\ref{fig_ratio}, upper-right panel). 

Thus, provided that an appropriate $\kappa_\mathrm{abs}$ is used, MBB fits
can retrieve the dust mass within at most 20\% of what more complex models
including a distribution of dust grains (and temperatures) can. This difference
is smaller than the current uncertainties in MW $\kappa_\mathrm{abs}$
models themselves. Finally, it is worth noting that for some objects SEDs 
are well fitted by neither a MBB, nor a DL07 model (D12).

\section{MBB dust masses vs $\beta$}
\label{sec_beta}

D12 also made MBB fits allowing $\beta$ to vary. For most of the objects, they 
find $\beta\approx1.5$. 
The analysis of Herschel colors of galaxies does suggest that $1\la \beta \la 2$ 
\citep[see e.g][]{BoselliA&A2012,AuldMNRAS2012}.
Keeping the same cross section normalization,
D12 find that dust masses can be severely underestimated.
Indeed, if I use the normalization of Eq.~\ref{fit_d03} together with $\beta=1.5$ in 
Eq.~\ref{eq_emy}, the median ratio between MBB and DL07 
masses reduces to 0.55, 0.51 and 0.54 for the three comparison in Fig.~\ref{fig_ratio} 
(see the lower panels, filled symbols). 
This is due to the change in the fitted $T_\mathrm{d}$, whose median value raises from 
$\approx$20K for $\beta=2.08$ and 21K for $\beta=1.91$ to $\approx$24 for $\beta=1.5$ 
(for a fixed normalization of $\kappa_\mathrm{abs}$, a smaller amount of hotter 
dust can reproduce the observed fluxes).

Despite it being common practice, it is however not justifyied to use 
$\kappa_\mathrm{abs}(\lambda_0)$ from a dust model (which has a proper
spectral index, close to $\beta=2$ for both models considered here)
and then assume a different spectral index. This inconsistency leads to the 
puzzling result that the dust mass estimate can vary with the wavelength chosen 
for the normalization: for the case of $\beta=1.5$, adopting the model cross section 
at 500$\mu$m results in dust masses that are 50\% to 30\% higher than those obtained 
with the 250$\mu$m normalization of Eq.~\ref{fit_d03} and \ref{fit_mc10}, respectively. 
The difference simply comes from
$(500/250)^{(\beta-1.5)}$, the fitted temperature being independent of the choice
for the normalization. A similar remark was made by \citet{SkibbaApJ2011}, which
fitted KINGFISH SEDs from Spitzer and SPIRE using $\beta=1.5$ and 
$\kappa_\mathrm{abs}(500\mu\mathrm{m})$ from D03. However, they inexplicably concluded 
that their normalization results in a mass that is a factor 3 higher than when 
$\kappa_\mathrm{abs}(250\mu\mathrm{m})$ is used.  The temperatures from \citet{SkibbaApJ2011}
are sistematically higher than those obtained here for $\beta=1.5$, and the dust masses, 
when corrected to the distances and normalization used here, are lower by a factor 2.7;
A similar discrepancy was also noted by \citet{GalametzMNRAS2012} and imputed to the use 
of $70\mu$m Spitzer data, which is contaminated by emission from stochastically
heated grains.

For a proper mass estimate, one would need
an absorption cross section derived consistently with the assumption 
made on $\beta$. This could come from a galaxy whose SED behaves as a MBB with the chosen $\beta$,
provided the dust mass can be derived with sufficient accuracy independently of the SED 
itself (for example using the mass of metals as a proxy, as done by \citealt{JamesMNRAS2002}); 
or from a dust model for the MW grains in which the materials cross sections follow 
the chosen spectral behavior (provided that those materials are able to
predict consistently both the MW extinction curve and the FIR/submm emission).

As a numerical experiment, I show here what a different choice of $\beta$ could imply in
the derivation of the absorption cross section from the MW SED. C11
derived the FIR/submm surface brightness for the Diffuse High Galactic Latitude medium
(DHGL, with latitude $|b|>15^\circ$) normalised to the hydrogen column density, $I_{\nu}/N_\mathrm{H}$. 
They provide Herschel surface brightnesses for the DHGL by convolving their COBE-FIRAS spectrum 
with the PACS (160$\mu$m) and SPIRE (extended emission) filter response functions.
They also provide the COBE-DIRBE $100\mu$m surface brigthness, which I converted to
PACS (100$\mu$m) by multiplying it by 1.1 to take into account the different filter
widths. This dataset is thus analogous to what I used in Sect.~\ref{sec_mass} to
derive dust masses.

For a dust grain of temperature $T_\mathrm{d}^\mathrm{MW}$ emitting as a MBB with a
power-law absorption cross section (i.e. the same assumptions as in Sect.~\ref{sec_mass}), 
the normalization of the cross section can be derived by fitting
\begin{equation}
\frac{I_{\nu}}{N_\mathrm{H}}=\frac{\tau_\mathrm{abs}}{N_\mathrm{H}} \; B_\nu(T_\mathrm{d}^\mathrm{MW}),
\label{eq_mw}
\end{equation}
where
\begin{equation}
\frac{\tau_\mathrm{abs}}{N_\mathrm{H}} = \kappa_\mathrm{abs} \; m_\mathrm{H} \; (D/G)_\mathrm{MW} = 
\frac{\tau_\mathrm{abs}(\lambda_0)}{N_\mathrm{H}}\times \left(\frac{\lambda_0}{\lambda}\right)^\beta.
\label{eq_t0}
\end{equation}
The absorption cross section per H atom, $\tau_\mathrm{abs}/N_\mathrm{H}$, can be 
converted into the cross section per unit dust mass of Eq.~\ref{eq_emy} by
dividing it for the hydrogen nucleon mass $m_\mathrm{H}$ and the MW dust-to-gas mass ratio
$(D/G)_\mathrm{MW}$. 

\begin{figure}
\centering
\includegraphics[width=\hsize]{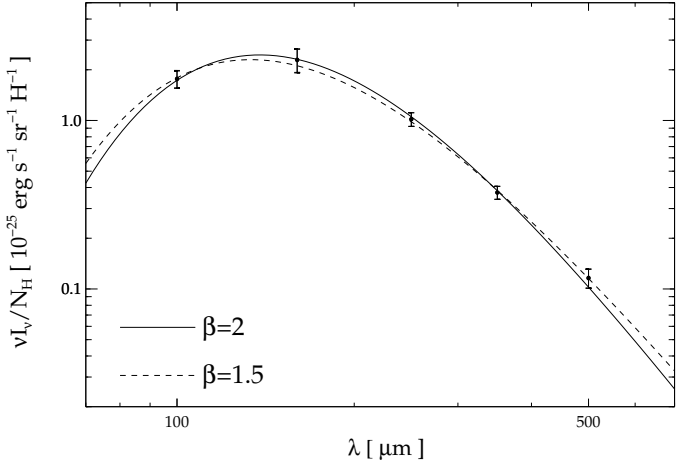}
\caption{MBB fits to the DHGL emission.}
\label{fig_mwfit}
\end{figure}

Using $\beta=1.91$, the DHGL emission can be fitted\footnote{I stress here that the aim of 
this numerical tests is only to show how the results are affected by the choice of $\beta$, 
and not to derive the spectral index of MW emission. Though $\beta$s between 1.5 and 2.1 
produce plausible fits in the limited spectral range considered in this work, the analysis 
of the full MW dust SED yields $\beta\approx1.8\pm0.2$ \citep{PlanckEarlyXXIV}.}
with $T_\mathrm{d}^\mathrm{MW}=17.8\pm0.4$K.
At 250$\mu$m, the absorption cross section is 
found to be $\tau_\mathrm{abs}(250\mu m)/N_\mathrm{H}=0.84\pm0.09 \times 10^{-25}$ cm$^2$ H$^{-1}$.
Assuming $(D/G)_\mathrm{MW}\approx0.01$, as can be estimated from the elemental depletion patterns
in the diffuse MW gas \citep{DraineBook2011}, it is $\kappa_\mathrm{abs}(250\mu m)
\approx 5.0 \; \mathrm{cm}^2 \; \mathrm{g}^{-1}$, a value close to that from C11
in Eq.~\ref{fit_mc10}: this is not surprising, since the dust model was made to fit the DHGL
data and has a similar dust-to-gas ratio, $(D/G)_\mathrm{MW}=0.0102$. For $\beta=2$, 
$T_\mathrm{d}^\mathrm{MW}=17.4\pm0.4$K and $\tau_\mathrm{abs}(250\mu m)/N_\mathrm{H}=0.91\pm0.10 
\times 10^{-25}$ cm$^2$ H$^{-1}$ (Fig.~\ref{fig_mwfit}), a
value close to the original determination of \citet{BoulangerA&A1996} on the COBE-FIRAS spectrum
(though C11 correct for the contribution of ionised gas to the hydrogen
column density, while \citeauthor{BoulangerA&A1996} only consider atomic gas). When $\beta=2.08$,
as for D03's model, $T_\mathrm{d}^\mathrm{MW}=17.1\pm0.4$K and 
$\tau_\mathrm{abs}(250\mu m)/N_\mathrm{H}=0.98\pm0.11 \times 10^{-25}$ cm$^2$ H$^{-1}$, which 
converts to $\kappa_\mathrm{abs}(250\mu m) \approx 5.9 \; \mathrm{cm}^2 \; \mathrm{g}^{-1}$, a 
value 45\% higher than that in Eq.~\ref{fit_d03}. 
The model could still be reconciled with the data,
if the ionised gas contributes more to the total hydrogen column density than
what assumed by C11. To be consistent with the mass determinations
done so far, I continue to use the absorption cross sections from Eqs.~\ref{fit_d03} and \ref{fit_mc10}
but scale them according to the ratios of $\tau_{\mathrm{abs}}/N_\mathrm{H}$ obtained from 
the DHGL with different $\beta$s.

Using $\beta=1.5$ the temperature rises to $T_\mathrm{d}^\mathrm{MW}=19.7\pm0.5$K
and the cross section reduces to $\tau_{\mathrm{abs}}(250\mu m)/N_\mathrm{H}=0.57\pm0.06 
\times 10^{-25}$ cm$^2$ H$^{-1}$, a factor 0.58 and 0.68 those for $\beta=2.08$ and 1.91, 
respectively. Obviously, the change of $\beta$ has the same effect on both the dust mass and cross 
section determination, since Eq.~\ref{eq_mbb} and \ref{eq_mw} are formally identical.

When the normalizations of Eq.~\ref{fit_d03} and \ref{fit_mc10} are corrected by those factors,
the dust masses obtained using $\beta=1.5$ become very close to those obtained using $\beta$ 
fitted to the dust models (see the open symbols in the lower panels of Fig.~\ref{fig_ratio}, 
with median mass ratios of 0.94, 0.88 and 0.8 for the three panels, from left to right).
Thus, as already noted in \citet{BianchiA&A1999}, the dust mass estimate does not change 
significantly with $\beta$ if both the absorption cross section and the dust mass are derived 
with the same 
$\beta$. Substituting Eq.~\ref{eq_emy}, \ref{eq_mw} and \ref{eq_t0} in Eq.~\ref{eq_mbb}, one finds 
that the dust mass of a galaxy depends on $\beta$ only through the ratio
\[
M_\mathrm{d}\sim \frac{B_\nu(T_\mathrm{d}^\mathrm{MW})}{B_\nu(T_\mathrm{d})}.
\]
For the wavelength and temperature ranges considered here, this ratio does not change significantly if
the temperatures are derived using $\beta=1.5$ or $\beta\approx2$.

\section{Conclusions}
\label{sec_conclu}

I have shown that, if the DL07 model of dust emission in galaxies is correct, a simple 
single temperature MBB fit to the SED for $\lambda\ge 100\mu$m can provide in a reliable 
way one of the parameters of the more complex approach, i.e. the dust mass. For this,
it is necessary to ensure the consistency in dust emission properties between the two
approaches: the absorption cross section used in the MBB needs to be the average over the grain size
distribution and composition of the dust model used in DL07. This consistency is broken
when a cross section normalization at a given wavelength is taken from, e.g., the D03
and C11 grain models (which have been derived with $\beta\approx2$, implicity coming from
the adopted material properties) and used in conjunction with $\beta\napprox 2$ 
in the MBB fits.

The need for consistency is illustrated by a numerical experiment: using either $\beta\approx2$
or 1.5 to derive the absorption cross section from the MW SED for $100\mu m \le \lambda\le 500\mu$m, and
the same $\beta$ to derive the dust mass from the SED of galaxies (both under the assumption
of a single temperature MBB) results in dust masses which are almost independent of
$\beta$. This might be true only for this limited range, which is still able to provide a
reliable fit of the MW SED. For $\beta$s outside this range, the normalization 
of the absorption cross section must come from other sources.

Finally, I remind the reader that I have only considered the SED for $\lambda\ge 100\mu$m.
Single temperature MBB fits to datasets including flux densities at shorter wavelengths 
(i.e.\ with a large contribution of non-equilibrium emission), as well as two temperatures 
MBB fits, might result in further biases in the derivation of the dust mass, in 
addition to those discussed here \citep[][Aniano et al.\ in preparation]{AnianoApJ2012}.

\begin{acknowledgements}
I thank Daniel Dale for providing me the parameters of the DL07 fits, Bruce Draine 
for stimulating discussions, Carlo Giovanardi and Leslie Hunt for useful comments.
\end{acknowledgements}

\bibliographystyle{aa}
\bibliography{/Users/sbianchi/Documents/tex/DUST}

\end{document}